# Resilience in Highways: Proposal of Roadway Redundancy Indicators and Application in Segments of the Brazilian Network

André Borgato Morelli

André Luiz Cunha

University of São Paulo

São Carlos School of Engineering

**ABSTRACT**

With the growing realization that transport systems must operate satisfactorily not only in typical situations, but also in adverse circumstances, ensuring redundancies in road systems has gained crucial importance. In this context, several methods have been proposed for measuring the vulnerabilities and resilience of transport systems. However, a simple metric to understand and quantify the degree of redundancy of a given road segment is still necessary, mainly to guide the responsible bodies regarding the need for intervention or special care with certain sections of the system. Thus, this paper proposes a redundancy indicator based on network analyses in the vicinity of an element. The proposed indicator was first calculated on nine application examples and then on a substantial sample of the Brazilian road network (~10% of segments). The results demonstrate that the indicator can satisfactorily describe the variety of cases in the Brazilian network, capturing cases where there is significant redundancy in the elements, as in some regions of the Southeast and South; or cases of very low redundancy, such as the sparse grid in the north of the country. It was also verified that the indicator has a particular sensitivity to parameters of the defined function, requiring further research for an acceptable calibration.

## 1. INTRODUCTION

In recent years, there has been growing research interest in understanding the operational behavior of transportation systems under adverse conditions, including situations such as flooding (Borowska-Stefańska et al., 2019; Chen et al., 2015; Kasmalkar et al., 2020), earthquakes (Kilanitis and Sextos, 2019; Wu et al., 2021), fuel supply crises (Azolin et al., 2020; Martins et al., 2019), and other events that can temporarily or permanently disrupt transportation networks (Berche et al., 2009; Ip and Wang, 2011; Mattsson and Jenelius, 2015). This research has gained significance due to the increasing frequency and severity of natural phenomena resulting from climate change, which can impact roadways through storms and landslides.

In this context, the importance of establishing redundancy within highway systems has become increasingly evident. Redundancy, in this context, refers to network elements that perform similar functions under normal conditions, such as offering alternative routes with close travel times for origin-destination pairs. When redundancy exists, other elements can step in as a substitute in cases where one section is damaged, blocked, or severely affected. The significance of redundancy becomes especially clear when aiming to enhance the resilience of the roadway system, as it can effectively reduce vulnerability to disruptions. This ensures the continuous flow of traffic even when components fail or are damaged, minimizing the disruptions caused by traffic interruptions and bolstering the overall ability of the network to withstand and adapt to adverse conditions (Mattsson and Jenelius, 2015).

Prior studies in the field of transportation system vulnerabilities have predominantly employed graph theory metrics to analyze transportation networks (Berche et al., 2009; Furno et al., 2018; King et al., 2020; Mattsson and Jenelius, 2015; Morelli and Cunha, 2021a; Reggiani et al., 2015). For instance, Berche et al. (2009) pioneered the evaluation of gradual element removal from a public transportation network using graph theory centralities. They identified betweenness centrality as a critical factor in assessing vulnerabilities. Also, García-Palomares et al. (2013) extended this perspective by assessing the resilience of a public transportation system (Madrid metro, Spain) and considering the number of affected users in their analysis.

Beyond graph theory measures, other studies have proposed various indicators for characterizing transportation network resilience, including indicators related to a city's connectivity within a road network (Ip and Wang, 2011) and network aspects like compactness and throughput (Zhang et al., 2015). King et al. (2020) conducted a comprehensive analysis of vulnerability in public transportation networks, combining graph theory metrics with network simulations to identify high and low-risk zones in Toronto, Canada.

The primary limitation of many of these studies lies in their focus on assessing if there is a route between origin-destination pairs, ignoring the increase in length/travel time. This approach is well-suited for urban environments, where detours typically result in only marginal increases in travel distances due to the higher density of urban road networks. Consequently, connectivity-related problems in urban settings primarily arise when significant impacts lead to the disconnection of several segments within the city.

Other studies have focused on parameters more directly related to transportation, such as total delay resulting from the deactivation of network elements in urban settings (Kasmalkar et al., 2020) and their impact on user accessibility (Borowska-Stefańska et al., 2019). Additionally, Cox et al. (2011) explored economic aspects associated with potential terrorist attacks on public transportation systems with a specific case study in London. While these analyses offer valuable insights and require precise data, they are typically designed to address more extensive impacts (involving multiple segments or entire regions) within urban environments. Consequently, they may be less effective in assessing smaller-scale impacts, such as the deactivation of individual segments, which may be imperceptible in the overall system unless they involve critical chokepoints like bridges.

Also, in rural networks, detours can substantially increase travel distances for local users, rendering routes impractical even when there is a viable alternative. Moreover, localized issues within interurban transportation systems, even if they do not significantly affect the overall network, can still substantially impact the immediate region, causing supply shortages and mobility challenges for entire cities.

Therefore, this research introduces a novel redundancy indicator for segments within the Brazilian National Road System (SNV in Portuguese) (DNIT, 2022) to identify vulnerabilities and critical points within the network. Distinguishing from previous studies, this indicator assesses the specific impact of an SNV segment on users who rely most directly on it, i.e., those who would need to detour from their routes in the absence of the segment. This approach allows for the characterization of the regional significance of a segment, even if its impact on the national network is limited. Furthermore, the assessment radius is confined to a predetermined range, reflecting the assumption that long-distance routes tend to deviate from their usual paths only when network issues are nearby.

The primary advantage of this novel indicator is its capacity to characterize redundancy levels on a segment-by-segment basis, thereby assessing the potential impact of deactivation due to natural disasters or related events. This framework facilitates the integration of resilience analysis into the roadway system's planning, operation, and maintenance. Moreover, the indicator's computational simplicity and relatively low complexity make it particularly suitable for regulatory agencies and transportation planners. Its ease of use also allows for evaluating multiple scenarios at a minimal computational cost, enabling the assessment of the impact of implementing new segments or planned maintenance.

## 2. PROPOSED METHOD

In this section, we propose redundancy indicators and describe the data used for evaluating and analyzing the sensitivity of the indicators in the National Road Network.

## 2.1. Definition of Alternative Routes

Characterizing the redundancy of a road infrastructure section involves measuring the number of alternative routes between a given origin-destination pair (Xu et al., 2021). However, the first challenge in defining a redundancy indicator in road systems lies in understanding what constitutes an alternative path. Geometrically, this problem results from two main factors:

1. Transportation networks are complex and relatively dense systems, such that removing a single element from the network rarely prevents users from traveling from one point in the network to any other. However, an alternative path can deviate so drastically from the original route that, in practical terms, it renders the route unviable, as shown in Fig. 1(a). In these cases, what is measured is not the network's ability to remain connected but its ability to offer acceptable alternative routes compared to the original route.

2. The magnitude of the deviation caused by a network failure depends directly on the origin and destination of the user in question, as shown in Fig. 1(b). While the deviation for one user may be acceptable, resulting in a slight increase in their route, it may include multiplying the initial distance by several times for another user[1].

---

[1] Note that what distinguishes the validity of routes for different origin-destination pairs in this proposition is determined by their final length. However, there is also the case of a route becoming impossible, i.e., the absence of an edge makes the path between $i$ and $j$ impossible. To standardize the distance criterion, it is considered that the length of an impossible path tends to infinity.

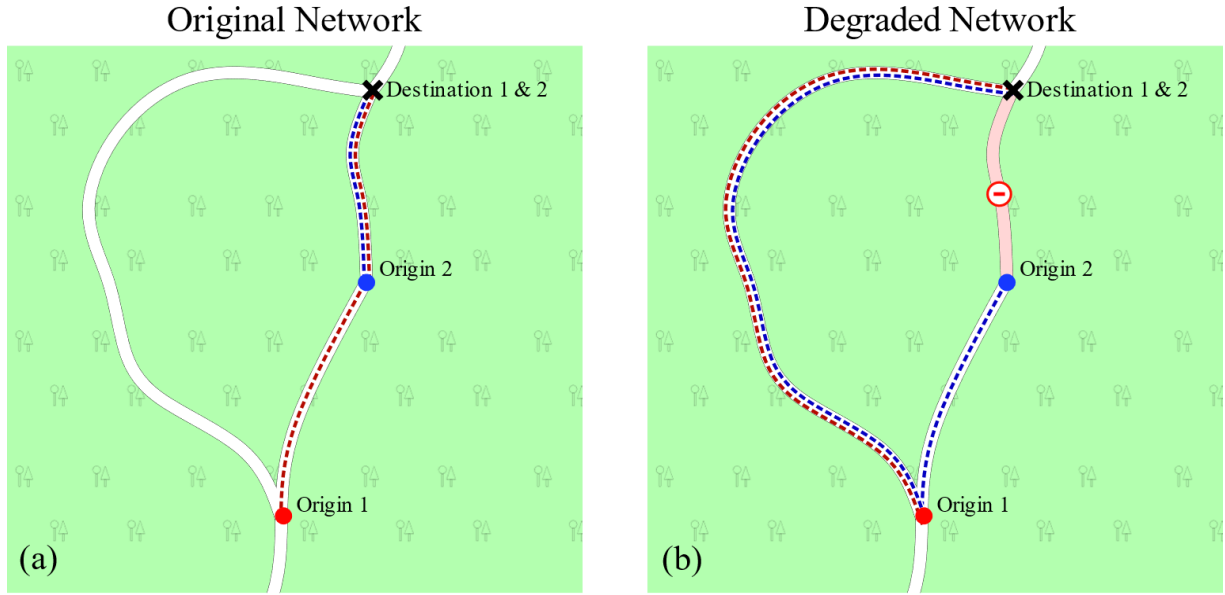

**Fig. 1:** Network without degradation (a) and network with one deactivated edge (b). While the path from origin 1 to the destination experiences a slight change of route, the path starting from origin 2 significantly increases its initial length.

Considering these two issues, we assume in this work that an element has low redundancy when, on average, it cannot provide a valid alternative to origin-destination pairs in its vicinity. Thus, the problem can be formulated as:

$$I_r(e) = \frac{1}{N} \sum_{i,j \in R;\ \sigma_{ij}(e) \geq 1} f(i,j) \qquad \text{Eq. 1}$$

where $I_r(e)$ is the redundancy indicator of the segment and the network; $N$ is the number of origin-destination pairs in the region of interest $R$ of the element; $\sigma_{ij}(e)$ is the number of shortest paths between $i$ and $j$ that pass through $e$; and $f(i,j)$ is the validity function for a route between origin $i$ and destination $j$. The definition $\sigma_{ij}(e) \geq 1$ ensures that only origin-destination pairs that depend on edge $e$ are considered in the validity analysis, considering only the impact on users who use the impacted infrastructure element. This criterion ensures that the evaluated impact refers only to users dependent on this segment to execute their shortest paths, unlike what is proposed in other studies that assess the impact even for users who did not originally depend on the segment in question (Berche et al., 2009; Morelli and Cunha, 2021a; Rodríguez-Núñez and García-Palomares, 2014).

The function $f(i,j)$ is the main element of this formulation since it is responsible for numerically representing how valid an alternative route is in the absence of element $e$. In the subsequent subsections, we define two strategies for defining this function: a conditional one, which defines only two levels for the function (0 for invalid routes and 1 for valid routes), and a continuous one in the range from 0 to 1 and reflects the efficiency of an alternative route in replacing the original one.

## 2.2. Conditional Validity Function

The first approach to defining $f(i,j)$ is to set an acceptable limit for the proportional increase in route length in the absence of a specific element. Thus, considering a threshold condition $k$ for the validity of a route:

$$f(i,j) = \begin{cases} 1 \ if \ \frac{d_{i,j}(e)}{d_{ij}} \leq k \\ 0 \ otherwise \end{cases} \qquad \text{Eq. 2}$$

Where $d_{ij}$ is the shortest distance in the network between points $i$ and $j$ in the unchanged network, and $d_{ij}(e)$ is the shortest distance in the network between points $i$ and $j$ with the removal of edge $e$. When a path becomes impossible in the absence of edge $e$, its distance $d_{ij}(e)$ is considered to tend to infinity, zeroing the function $f(i,j)$.

The advantage of defining the validity function in this way is the ease of interpreting the obtained result. By associating Eq. 2 with Eq. 1, the redundancy index reflects the proportion of valid paths after network degradation. In this case, $I_r = 0.5$ means that half of the routes between accesses around the degraded element become unviable while the rest remain viable.

The disadvantage of this approach lies in the dependence on the definition of the parameter $k$. If one opts for a definition of $k$ based on observation, research involving users who can indicate a suitable average value for this parameter will be necessary. However, it is also likely that tolerance for deviations varies from region to region in the country. On the other hand, a governmental agency can define this value normatively – based on a value judgment – and define the deviation that would be reasonable for a user. In this work, we do not aim for any of these objectives, merely conducting a sensitivity analysis for the parameter on Brazilian highways as a first step towards a possible method implementation.

## 2.3. Continuous Validity Function

In the event of degradation in the network, there is a tendency for distances between origins and destinations to increase or, if there are multiple equally good alternatives, to maintain the same. In this case, the ratio of the original length of the shortest path between two points to its length in the degraded network naturally results in a number between zero and one, being zero when the degraded length tends to infinity (unviable route) and one when there is another path as good as the original.

$$f(i,j) = \frac{d_{ij}}{d_{ij}(e)} \qquad \text{Eq. 3}$$

This measure is called alternative efficiency and was used in an urban context (Morelli and Cunha, 2021b) to measure the impact of floods on the transportation system. The measure is defined as the inverse of the relative increase in distance to maintain the range between [0, 1] and to avoid sums with values tending to infinity in Eq. 1. The advantage of using this measure compared to the conditional application (item 2.2) is that it does not depend on any calibration parameter. However, this measure has the disadvantage of being more challenging to interpret. For instance, $I_r(e) = 0.5$, it means that there was an average 50% reduction in the efficiency of routes, which could be caused: by the minimum path of all original routes doubling, by half of the routes becoming unviable without affecting the other half, or some scenario in between. It is impossible to distinguish between the cases merely observing the indicators.

**2.4. Region of Interest**

In this work, we choose to evaluate alternative routes only within a region delimited by a certain radius around the geometric center of the assessed infrastructure element, as shown in Fig. 2. This choice is driven by three main reasons: (1) shorter-distance trips tend to be more frequent than longer ones in general; (2) longer trips, covering thousands of kilometers, tend to have routes with established intermediate stopping points along the way due to habit or contract, which results in deviations only occurring in the immediate region of the blockage occurrence; and (3) the computational cost of evaluating all shortest paths in a set of $n$ points requires processing $n(n-1)$ shortest paths. Therefore, evaluating all points in the network rather than just in the region of interest may increase computational complexity to render the analysis unfeasible on personal computers.

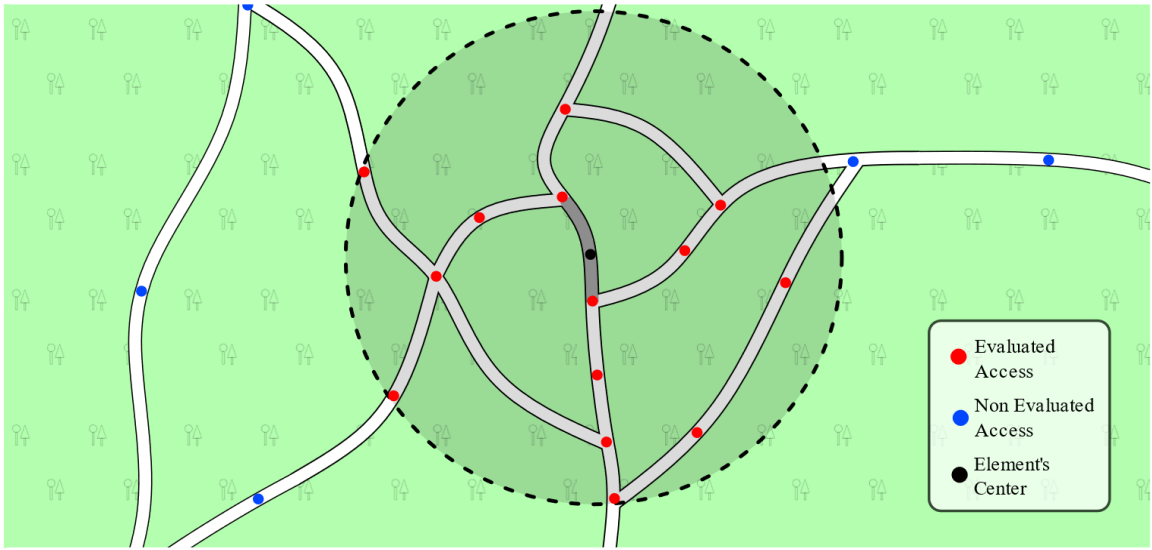

**Fig. 2:** Region of interest for redundancy analysis of a road network element.

The origin-destination points $i$ and $j$ are defined only at the accesses within the region of interest, making the computation of deviations feasible while avoiding an excessive number of paths to be evaluated. We emphasize that although the nodes outside this region are not used as origins, they are not removed from the original graph. They can still be used for the shortest path between two points in the region of interest, thus keeping the original graph topologically intact.

Furthermore, defining the region of interest based on a radius around the element adds one more calibration parameter to the model. Therefore, this work presents a sensitivity analysis of this parameter but leaves it to future research to determine the appropriate value for applications in Brazilian territory.

### 2.5. Database

In this work, we utilized the SNV provided in geographic file format (DNIT, 2022). We transformed this system into a bidirectional graph in Python for route and shortest path calculations. However, the provided geographic file had various internal inconsistencies, such as misalignments and disconnections of edges, which hindered its use for route analysis. Thus, the database underwent initial processing to address these inconsistencies programmatically. A more detailed description of the preprocessing of the SNV can be found in Appendix A of the supplementary material.

# 3. APPLICATION EXAMPLES

This chapter presents practical applications of redundancy calculation in select segments of the SNV network, along with a brief analysis of how the indicators vary with calibration parameters. We have chosen nine segments within the network, encompassing elements with varying degrees of redundancy, from those with clear alternative routes nearby to those with little or no redundancy. Table 1 lists these segments in order from the most redundant to the least.

To analyze how the radius of the region of interest and parameter $k$ affect the indicator in different scenarios, we tested these nine scenarios with radii ranging from 25 km to 500 km and $k$ values between 1.0 (100%) and 6.0 (600%). The indicator is computed multiple times for each element in these segments, considering various parameter combinations. The computation time for each scenario depends on the number of origins/destinations in the vicinity, which is proportional to the square of the radius of the region of interest and the density of access points in the region.

**Table 1:** Selected segments as application examples

| ID | Segment's code | State | Roadway | Sorroundings [20km] | Sorroundings [300km] |
|---|---|---|---|---|---|
| 1 | 153BPR1470 | PR | BR 153 | 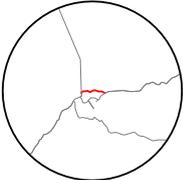 | 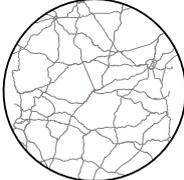 |
| 2 | 265BMG0170 | MG | BR 265 | 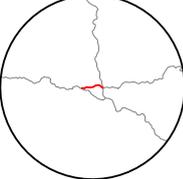 | 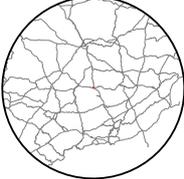 |
| 3 | 050BSP0550 | SP | BR 101 | 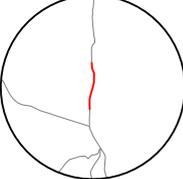 | 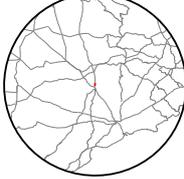 |
| 4 | 230BPB0510 | PB | BR 230 | 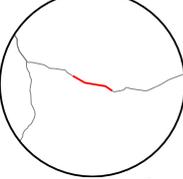 | 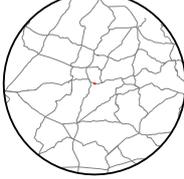 |
| 5 | 101BSP3710 | SP | BR 101 | 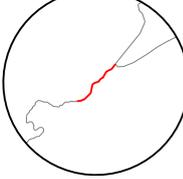 | 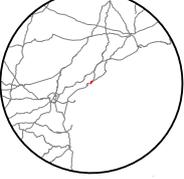 |
| 6 | 352BGO0110 | GO | BR 352 | 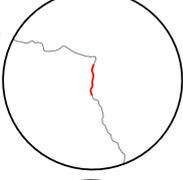 | 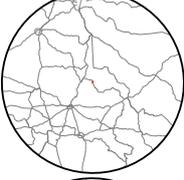 |
| 7 | 259BMG0250 | MG | BR 259 | 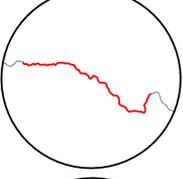 | 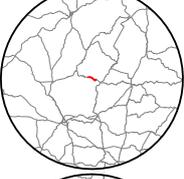 |
| 8 | 010BTO0310 | TO | BR 010 | 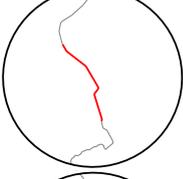 | 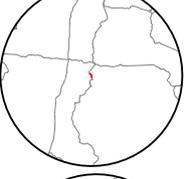 |
| 9 | 499BMG0020 | MG | BR 499 | 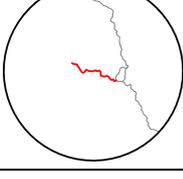 | 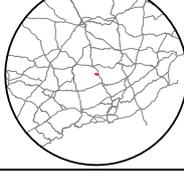 |

## 3.1. Conditional Validity Function

Figure 3 illustrates the behavior of the nine segments relative to the radius and the $k$ parameter variation. The general trend displayed by the elements is that the redundancy indicator increases with an increase in the radius, as expected since greater distances between the origins and destinations of the degraded element tend to lead to the existence of alternative routes of similar length. The indicator also tends to increase with higher tolerance values ($k$). An exception is segment 9 (499BMG0020) on BR 499 where, being a dead-end road, all paths passing through it necessarily become impossible when the road is blocked since the road is the sole viable route, resulting in the indicator assuming a value of zero for all instances.

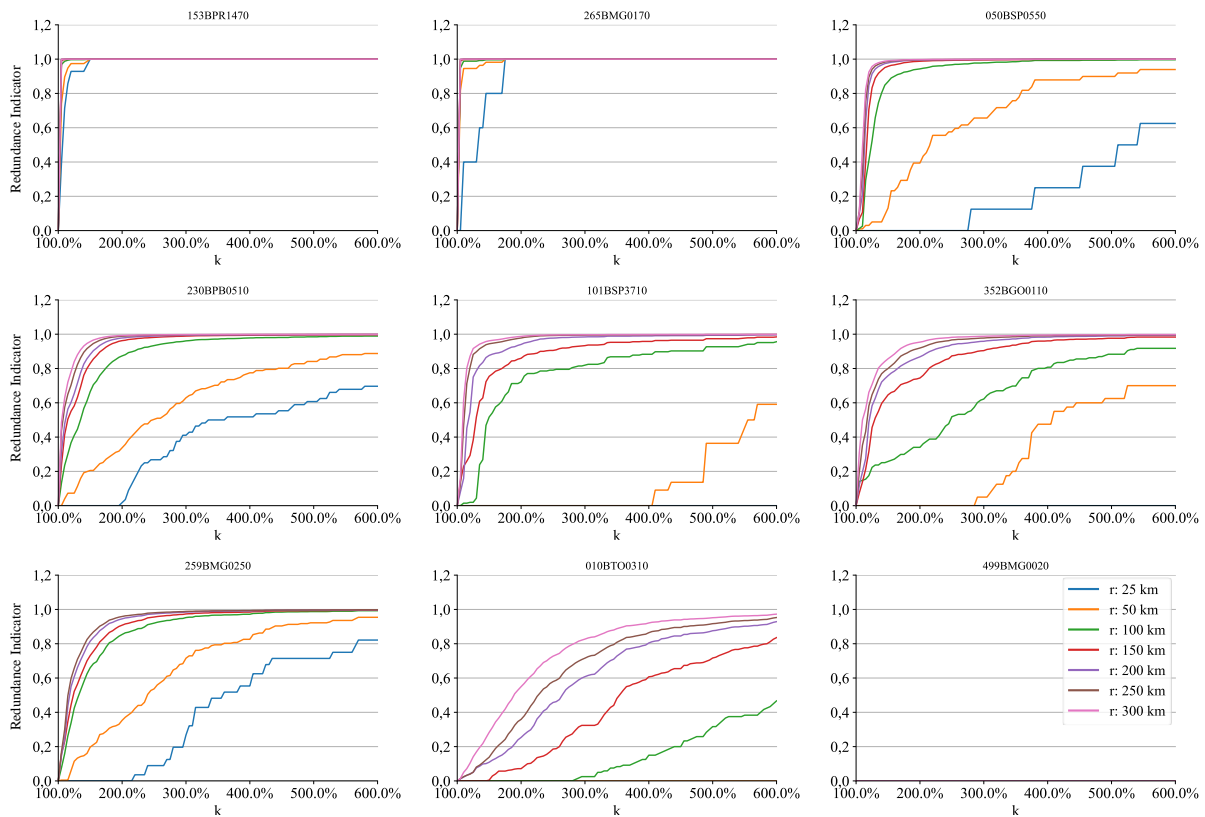

**Fig. 3:** Redundancy indicator with conditional validity function according to region of interest radius and user tolerance ($\boldsymbol{k}$)

It is also noticeable that small radii (between 25 and 50 km) tend to result in null indicators unless tolerance is relatively high. This result poses a challenge for the indicator since, in such cases, it is difficult to distinguish between a route without any alternatives and one with a few. On the other hand, as radii exceed 200 km, the indicators exceed 0.8 even for less redundant routes and lower tolerances. This suggests that the variance of results for excessive radii decreases, potentially complicating the application of the indicator. Larger tolerances also have a similar effect of reducing variance, albeit to a lesser extent. Thus, practical applications of the indicator require a reasonable radius value (between 100 km and 200 km) and $k$ value (between 1.5 and 3.0) to capture the redundancy effect with relatively large variance among elements. These effects are further examined with a larger sample in Section 4.

## 3.2. Continuous Validity Function

Figure 4 depicts how the indicator for each route varies with changing radius. Noticeably, the difference between the sampled road segments is reduced with larger radii. One difference between this sort of function and the conditional is that, except for segment 9 (499BMG0020), the indicator never reaches zero, regardless of how close it approaches it. Moreover, a route with marginally better alternatives than another will always have a higher indicator, no matter how minor the differences are, as the validity function assumes continuous values. However, as previously mentioned, there is the issue of needing more clarification on the results.

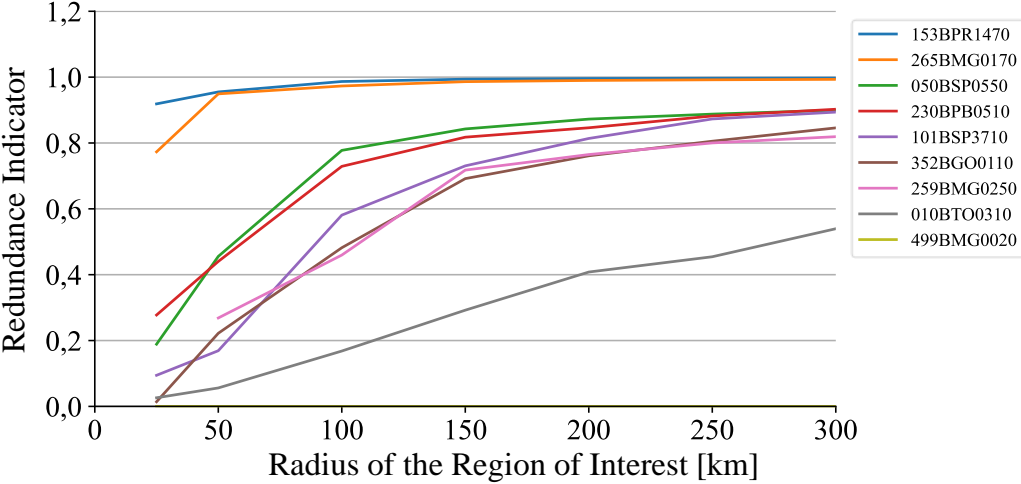

Fig. 4: Redundancy Indicator with Continuous Validity Function According to Region of Interest Radius

Regardless of the validity function used, the key takeaway is that the indicator can capture redundancy in locations where one would intuitively expect it. For example, segments 1 (153BPR1470) and 2 (265BMG0170), visually presenting a clear alternative nearby, tend to have a high indicator even for smaller radii and tolerances. In contrast, segments that lack nearby alternative connections due to their location in regions with lower connection density, such as 7 (259BMG0250) and 8 (010BTO0310), exhibit a trend towards lower indicators.

These results exemplify the utility of the indices in characterizing the redundancy of elements with a certain degree of mathematical rigor, facilitating the establishment of standards that consider route vulnerability in prioritizing environmental robustness or the implementation of new segments.

## 4. RESULTS FROM SAMPLE

In this section, we present an in-depth analysis with an increased number of segments to provide a clearer insight into the distribution trends of indicators across various radius and tolerance values. To achieve this, we employed a random sample comprising approximately 10% of all segments within the national territory (984 segments). Figure 5 displays the distributions of the indicator ($I_r(e)$) using violin plots (left) and cumulative distributions (right) for different radii and tolerances while considering a conditional validity function.

Radii of 25 km and 50 km tend to disproportionately concentrate indicators in the lower end of the spectrum, with most results indicating a zero value for redundancy. This outcome was anticipated, especially for the 25 km radius, partly because some segments in the network exceed the assessed diameter, resulting in no origin-destination pairs being computed in such cases. Conversely, a radius of 200 km tends to concentrate results in the upper spectrum, particularly in cases of higher tolerance. Figure 6 illustrates the identical distributions for the continuous validity function, showing similar overall behavior, with radii of 25 km and 50 km concentrating values close to zero. In comparision, the 200 km radius concentrates values near 1.0.

In this context, the combination that yields the most dispersed distributions is a radius of 100 km with a tolerance between 150% and 200%. This observation is supported by the dispersion statistics presented in Table 2, indicating that both variance and the distance between the second and third quartiles are higher in these cases. However, using the combination with the highest dispersion as a criterion may be better in practice. This assumption can only be confirmed through field research on user tolerance and the typical radius at which users tend to deviate from the given route in the event of a closure. In the absence of such studies, it seems reasonable to consider that the combination offering one of the highest variances among the results is better at distinguishing more redundant elements from the more vulnerable ones.

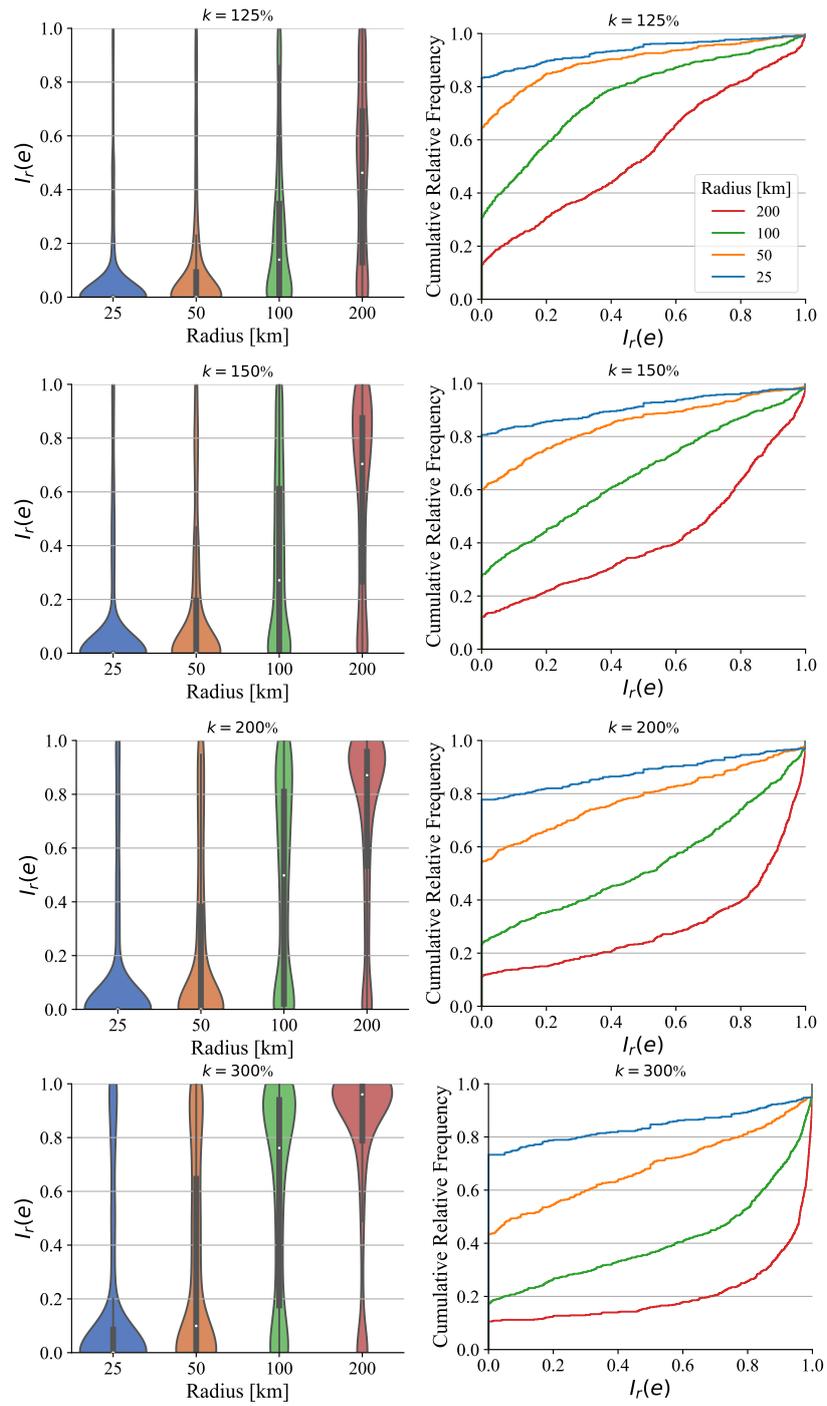

**Fig. 5:** Distribution of the redundancy indicator in violin plots (left) and cumulative relative frequencies (right) for the conditional validity function under various radius and $k$ scenarios.

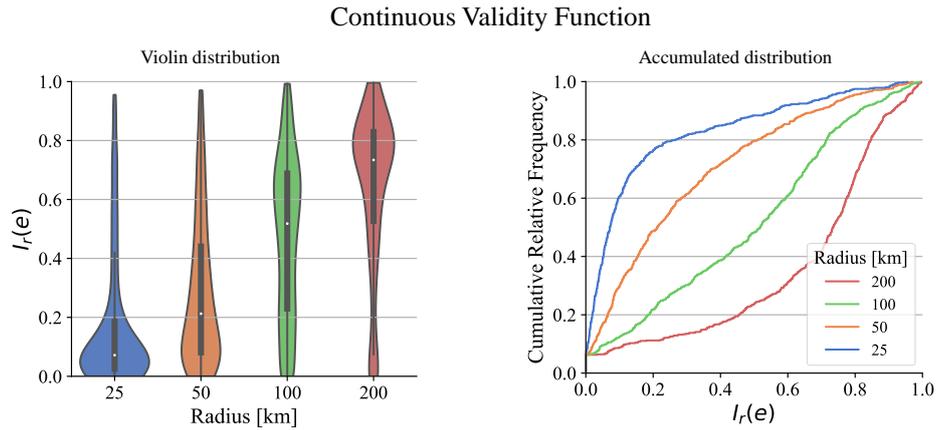

**Fig. 6:** Distribution of the redundancy indicator in violin plots (left) and cumulative relative frequencies (right) for the continuous validity function under various radius scenarios.

Figure 7 provides a map showing access density (a) and average road infrastructure redundancy (b and c; radius=100 km and k=200%) distributions by Intermediate Geographical Region, a territorial subdivision defined by the Brazilian Institute of Geography and Statistics (IBGE). Histograms of the distributions are displayed below the respective maps. There is a clear trend towards higher indicator values in regions with greater access density, as seen in the Southeast and South regions. This trend is supported by Figure 8, which shows a strong correlation between the logarithm of density and both indicators.

The strong correlation between the indicator and the number of accesses is expected if the indicator effectively captures redundancies in the system. In this case, regions with more accesses tend to offer a more significant number and quality alternatives, increasing redundancy on average, depending on other network topology parameters. This result further validates the quality of the proposed indicators.

Table 2: Sample statistics by validity function, radius, and $k$.

| Validity function | Radius [km] | $k$ | Redundancy Indicator ($I_r(e)$) | | | | | |
|---|---|---|---|---|---|---|---|---|
| | | | Mean | Variance | 25% | 50% | 75% | Q3-Q1 |
| Conditional | 25 | 1,25 | 0,064 | 0,034 | 0,000 | 0,000 | 0,000 | 0,000 |
| Conditional | 25 | 1,50 | 0,091 | 0,052 | 0,000 | 0,000 | 0,000 | 0,000 |
| Conditional | 25 | 2,00 | 0,121 | 0,072 | 0,000 | 0,000 | 0,000 | 0,000 |
| Conditional | 25 | 3,00 | 0,160 | 0,101 | 0,000 | 0,000 | 0,086 | 0,086 |
| Conditional | 50 | 1,25 | 0,103 | 0,048 | 0,000 | 0,000 | 0,093 | 0,093 |
| Conditional | 50 | 1,50 | 0,151 | 0,070 | 0,000 | 0,000 | 0,196 | 0,196 |
| Conditional | 50 | 2,00 | 0,217 | 0,101 | 0,000 | 0,000 | 0,383 | 0,383 |
| Conditional | 50 | 3,00 | 0,314 | 0,137 | 0,000 | 0,100 | 0,647 | 0,647 |
| Conditional | 100 | 1,25 | 0,233 | 0,078 | 0,000 | 0,140 | 0,348 | 0,348 |
| Conditional | 100 | 1,50 | 0,341 | 0,109 | 0,000 | 0,271 | 0,612 | 0,612 |
| Conditional | 100 | 2,00 | 0,463 | 0,137 | 0,021 | 0,498 | 0,810 | 0,789 |
| Conditional | 100 | 3,00 | 0,593 | 0,147 | 0,176 | 0,761 | 0,940 | 0,764 |
| Conditional | 200 | 1,25 | 0,445 | 0,105 | 0,129 | 0,463 | 0,691 | 0,562 |
| Conditional | 200 | 1,50 | 0,584 | 0,119 | 0,266 | 0,704 | 0,875 | 0,608 |
| Conditional | 200 | 2,00 | 0,703 | 0,118 | 0,533 | 0,871 | 0,958 | 0,425 |
| Conditional | 200 | 3,00 | 0,796 | 0,105 | 0,788 | 0,960 | 0,989 | 0,201 |
| Continuous | 25 | - | 0,160 | 0,101 | 0,000 | 0,000 | 0,086 | 0,086 |
| Continuous | 50 | - | 0,314 | 0,137 | 0,000 | 0,100 | 0,647 | 0,647 |
| Continuous | 100 | - | 0,593 | 0,147 | 0,176 | 0,761 | 0,940 | 0,764 |
| Continuous | 200 | - | 0,796 | 0,105 | 0,788 | 0,960 | 0,989 | 0,201 |

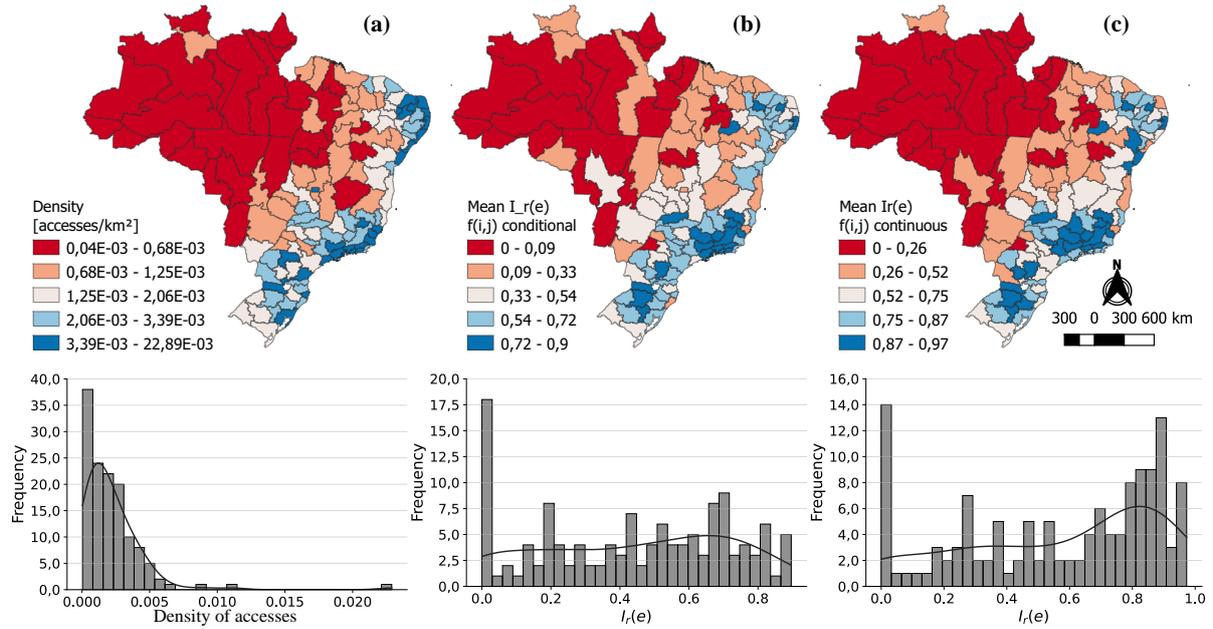

**Fig. 7:** Geographic distribution (above) and histograms (below) of access density and average indicators by Intermediate Geographical Region.

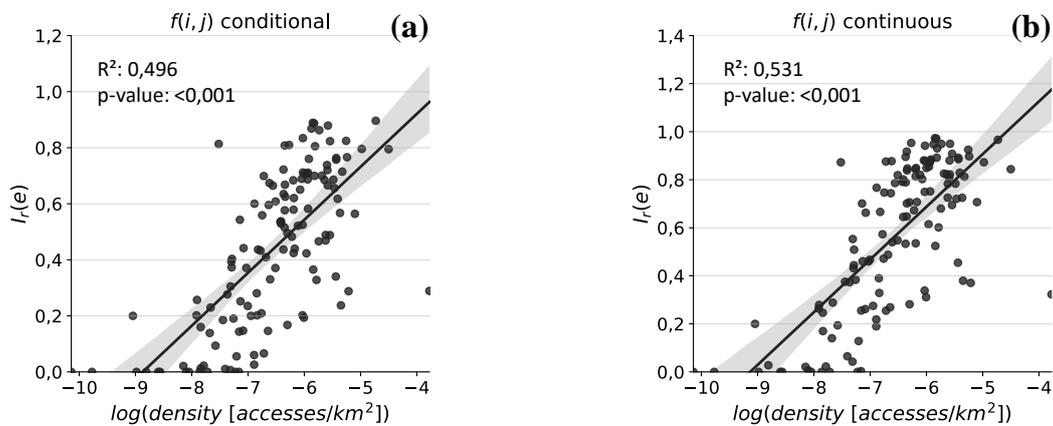

**Fig. 8:** Relationship between the logarithm of density and the redundancy indicator with conditional function (a) and continuous function (b) with a linear trend curve and 95% confidence interval.

## 5. CONCLUSIONS

The objective of this study was to propose a redundancy indicator to guide and prioritize investment in the Brazilian road network, aiming to enhance the resilience of the system in the face of disruptions. The developed metric is based on the number of valid alternative paths following the blockage of a road segment within the network. The validity of these alternative paths is conditioned by a validity function, which in this paper was defined as a conditional approach where an alternative route is considered valid if it satisfies a specific condition and a continuous approach where a continuous efficiency function determines route validity.

The developed metrics have proven effective in capturing the redundancy of a road segment, as demonstrated through nine practical applications and a brief spatial analysis. The spatial analysis revealed that the redundancy of a segment is proportional to the density of access points in its vicinity, aligning with intuitive expectations for a metric designed to measure network redundancy.

The method involves one to two calibration parameters, depending on the chosen validity function (radius of the region of interest and user tolerance). It was demonstrated that parameter ranges tend to increase result variability on a national scale (e.g., a radius of 100 km and tolerance between 150% and 200%). However, future studies can fine-tune these values based on user behavior to achieve a more accurate system evaluation.

Additionally, we acknowledge that road network resilience involves more than just network structure; it also requires ensuring that alternative routes can handle diverted traffic. In future studies, we plan to expand the method to account for traffic diversion effects on alternative route capacity, assessing both spatial and temporal disruptions. This analysis could provide a more realistic evaluation of system redundancy, but it will demand more extensive data, particularly regarding traffic patterns. Consequently, its widespread application across the national territory is expected to be more challenging than the indicators developed in this study.

**Funding**

This study was financed in part by the Coordenação de Aperfeiçoamento de Pessoal de Nível Superior - Brasil (CAPES) - Finance Code 001. This paper was also financed by the National Coucil for Scientific and Technologic Development (CNPq) – Grant number 311964/2022-2.

# APPENDIX A

In this section, we provide a detailed account of the preprocessing steps applied to the geographical dataset of the National Road Network System (SNV). The supplied geographical dataset exhibits two distinct types of inconsistencies that impact its suitability for route analysis: disconnection and misalignment of edges. Examples of both issues are depicted in Figure I, illustrating a set of disconnected edges (a1) and a set of misaligned edges (b1).

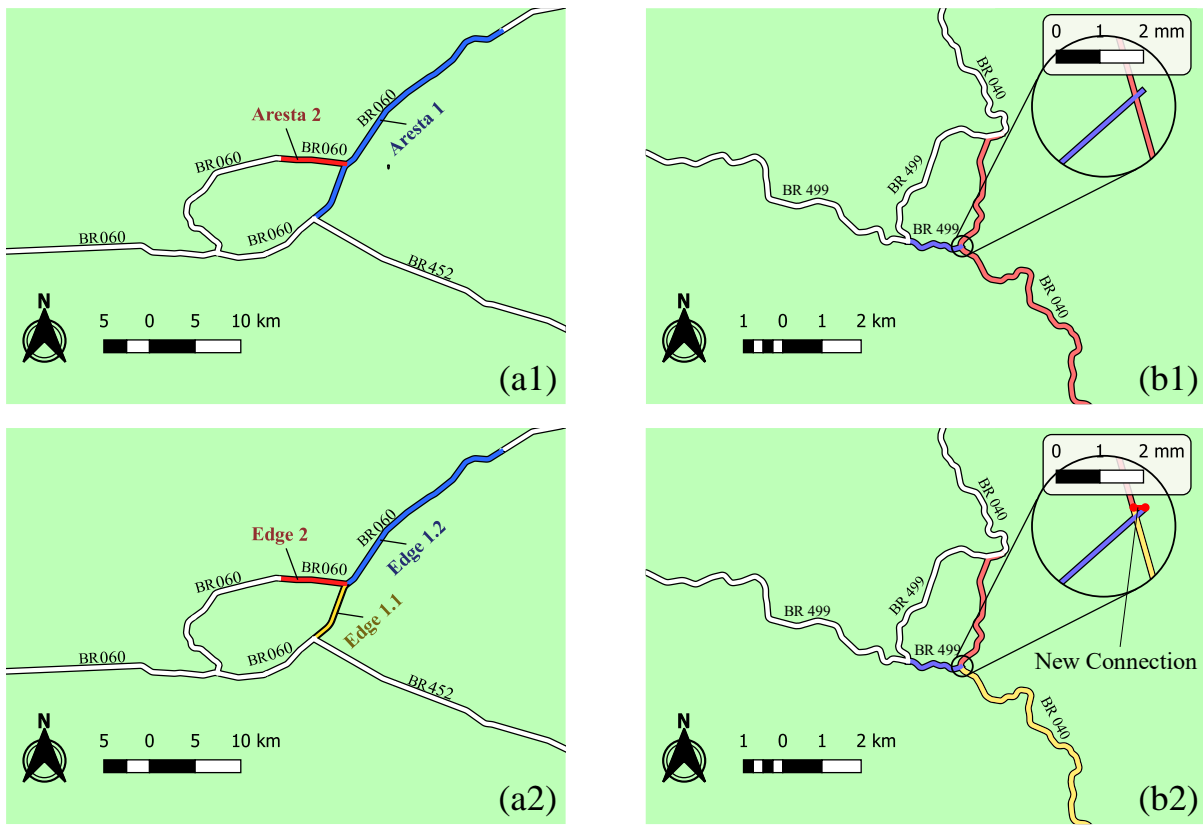

**Fig.I:** Disconnected edges (a1); Misaligned edges (b1); Disconnected edges after preprocessing (a2); Misaligned edges after preprocessing (b2).

The disconnection problem arises during the creation of the representative graph of the network. As per the graph's definition, a direct connection between two edges can only exist if their terminal nodes are interconnected. In the scenario presented in Figure I (a1), edge 1 lacks an endpoint at a node of edge 2. Consequently, for routing purposes, it becomes impossible to traverse from edge 1 to edge 2, and vice versa. These issues are remedied by segmenting edge 1 to establish a connection, as demonstrated in Figure A1(a2).

In the case of misalignment between two edges, one edge does not precisely terminate along the path of the other, resulting in an inherent disconnection issue, as exhibited in Figure I (b1). To address this problem, the misaligned edge is segmented at the point along its trajectory that is nearest to the misaligned node. Subsequently, the misaligned node is connected to the newly created node, establishing a topological connection, as illustrated in Figure A1(b2).